\documentclass[10pt,aps,prd,nofootinbib,superscriptaddress,
  twocolumn,preprintnumbers,balancelastpage]{revtex4}
\PassOptionsToPackage{slash}{url}
\usepackage{graphicx,epsfig,psfrag,amssymb,hyperref}
\usepackage{multirow}
\usepackage{color,graphicx,epsfig,psfrag,amsmath,empheq}
\usepackage{bm}
\usepackage{mathrsfs,amsfonts,color}
\usepackage{slashed}
\usepackage{hepunits}
\usepackage{enumitem}
\usepackage{booktabs}
\usepackage{xspace}
\usepackage[utf8]{inputenc}
\usepackage{braket}

\AtBeginDocument{
\heavyrulewidth=.08em
\lightrulewidth=.05em
\cmidrulewidth=.03em
\belowrulesep=.65ex
\belowbottomsep=0pt
\aboverulesep=.4ex
\abovetopsep=0pt
\cmidrulesep=\doublerulesep
\cmidrulekern=.5em
\defaultaddspace=.5em
}

\pdfoutput=1
\newcommand{\ev}{\text{eV}}

\newcommand{\gev}{\text{GeV}}

\newcommand{\Eqref}[1]{Equation~(\ref{eq:#1})}

\renewcommand{\eqref}[1]{Eq.~(\ref{eq:#1})}
\newcommand{\eqsref}[2]{Eqs.~(\ref{eq:#1}) and (\ref{eq:#2})}
\newcommand{\secref}[1]{Sec.~\ref{sec:#1}}

\newcommand{\figref}[1]{Fig.~\ref{fig:#1}}

\newcommand{\tableref}[1]{Table~\ref{table:#1}}

\newcommand{\appref}[1]{Sec.~\ref{sec:#1}}

\newcommand{\beq}{\begin{equation}}
\newcommand{\eeq}{\end{equation}}
\newcommand{\beqa}{\begin{eqnarray}}
\newcommand{\eeqa}{\end{eqnarray}}
\newcommand{\nn}{\nonumber}

\newcommand{\ILag}[3]{{\cal I}_{#1,#2}(#3) }

\newcommand{\qmax}{  q_{\rm{max}}  }
\newcommand{\gsf}[1]{  \Gamma^{\rm sf}_{#1}  }

\begin{document}

\title{
Synchrotron-Like Radiation Beyond The Standard Model I: \\
Hunting for new physics with the Sokolov-Ternov effect}


\author{Iftah Galon}
\email{iftah.galon@physics.rutgers.edu}
\affiliation{New High Energy Theory Center \\
Rutgers, The State University of New Jersey \\
Piscataway, New Jersey 08854-8019, USA}

\begin{abstract}
Electron and positron beams in storage-rings self-polarize by emitting spin-flipping synchrotron radiation. If new ultralight particles couple to $e^\pm$, their emission in synchrotron-like radiation would modify the characteristic self-polarization time.
We calculate the rate of spin-flipping synchrotron-like radiation in several simplified models, and find that the largest contribution is for an axial-vector. 
We use polarization time measurements from the Swiss-Light-Source, and SPEAR3 to set new strong limits on ultralight axial-vectors coupled to $e^\pm$.
\end{abstract}

\vskip1cm

\maketitle

\section{Introduction}
\label{sec:intro}

Despite its success in describing experimental results up to the ${\rm TeV}$-scale, the standard model of particle physic (SM) cannot be a complete theory of Nature.
Primary deficiencies of the SM include: the lack of viable dark matter (DM) candidates, an explanation for neutrino oscillations, and an account for the Universe's matter/anti-matter asymmetry. Other issues include the hierarchy problem, the flavor puzzle, and the strong CP puzzle.
Theoretical extensions of the SM typically address a subset of these by introducing new degrees of freedom which couple to the SM particles. In such theories, DM candidates characteristically poses interactions, which make them testable in terrestrial experiments, and in non-gravitational astrophysical observations.
An interesting class of such new physics (NP) models is that of ultralight DM, with masses lower than $\sim10~\rm{eV}$~\cite{Essig:2013lka, Battaglieri:2017aum, Redondo:2010dp}. A yet unexplored avenue is to directly produce such particles
in storage-rings.

In storage-rings, an array of electric and magnetic field configurations
guides bunches of charged particles in orbits, causing these particles to emit synchrotron radiation. If these particles couple to new physics degrees of freedom, it is conceivable that the latter could also be emitted in {\it synchrotron-like} radiation. 
Synchrotron radiation has been widely studied both theoretically and experimentally (see~\cite{Sokolov:1986nk, Wiedemann2003, hofmann_2004} for reviews). 
Some characteristic observables, such as the power spectrum, are understood classically~\cite{Schwinger:1949ym}, and quantum mechanical effects modify them negligibly~\cite{Schwinger1954} for typical storage-ring energies~\cite{Chao:1981it}.
Nonetheless, particle beams in storage-rings exhibit important effects which are due to the quantum mechanical nature of synchrotron radiation. 
These effects can therefore constitute a precision test of Quantum Electrodynamics (QED).\footnote{
synchrotron-like production can also act as a DM-source~\cite{Galon_WIP}.
}

One such effect is the radiative polarization of electron or positron beams known as the Sokolov-Ternov (ST) effect~\cite{Sokolov:1963zn}. In a storage-ring, the synchrotron induced spin-flip transition rates are asymmetric between the two spin-states. Consequently, polarization builds up over a characteristic (machine-dependent) time scale, up to an asymptotic value. The effect has been extensively verified (see \cite{Shatunov:2006ng, Mane:2005xh, Gianfelice-Wendt:2018mcx} for reviews) using beam polarimetry techniques~\cite{Aulenbacher:2018weg}. Uses of the effect include beam energy measurements~\cite{Fischer:1979as}, and the production of polarized-beams in high-energy colliders like LEP~\cite{Knudsen:1991cu, Alexander:1988mv}, and HERA~\cite{Barber:1992fc, Sobloher:2012py}.

If new particles couple to electrons, their synchrotron-like radiation modifies the spin-flip transition rates. The aforementioned measurements therefore constrain the parameter-space of new-physics models which couple new light particles to electrons.

This work explores these constraints.
A basic review of the ST-effect is presented in \secref{ST_basics}, and the modifications for real storage-rings are elaborated in \secref{ST_real}.
\secref{new_physics} discusses the effects of new physics on polarization, describes a set of simplified models, and for each model,  presents the ``massless''-limit result for the spin-flip transition-rate.
The available measurements used in this work are presented in \secref{available_measurements} and the derived limits are given in \secref{limits}. Concluding remarks are given in \secref{conclusions}. 
The heavy lifting is reserved to the supplemental materials which include a short introduction to spin-dynamics and polarization in accelerator physics, as well as the technical details of the formalism, approximation methods, and the full results of the calculations.

\section{The Sokolov-Ternov Effect - Radiative Self Polarization of Charged Particles in storage-rings}
\label{sec:ST_basics}

In a storage ring with a uniform and constant background magnetic field, relativistic electrons and positrons emit synchrotron radiation which either flips or preserves their spin state.
Due to the magnetic field, the spin-flip transition-rates,
\beq
\label{eq:ST_rates}
\gsf{\gamma}(u)  \equiv
\Gamma_H(e^-(u) \to e^-(-u) \gamma) \, ,
\eeq
depend on the fermion spin state, $u=+,\, -$ for ``up`` and ``down'' along the magnetic field direction respectively ($\hat z$).
The polarization of the beam is the average
\beq
{\cal P} = \frac1{N}\sum_{\rm i=1}^N\braket{\vec S_i \cdot \hat z} = 
\frac{N_+\,-\, N_-}{N_+\,+\, N_-}
\eeq
where the sum runs over the $N$ particle spins, and $N_\pm$ is the number of particles in each respective spin state ($N = N_+ + N_-$).
The state population is governed by the rates in \eqref{ST_rates} using a transport equation
\beq
\frac{d}{dt}
\begin{pmatrix}
N_+ \\ N_-
\end{pmatrix}
=
\begin{pmatrix}
-\Gamma^{\rm sf}_{\gamma}(+) & \Gamma^{\rm sf}_{\gamma}(-) \\
\Gamma^{\rm sf}_{\gamma}(+) & -\Gamma^{\rm sf}_{\gamma}(-)
\end{pmatrix}
\begin{pmatrix}
N_+ \\ N_-
\end{pmatrix}
\,.
\eeq
For an initially unpolarized beam this leads to
\beq
\label{eq:pol_evolution}
{\cal P}(t) = {\cal P}_{eq}\left(1 - e^{- t / \tau_{p} }\right)\,.
\eeq
where the equilibrium polarization and the characteristic polarization time are respectively
\beq
\label{eq:pol_obs_def}
{\cal P}_{eq} = \frac{\Gamma^{\rm sf}_{\gamma}(+)\, -\, \Gamma^{\rm sf}_{\gamma}(-)}
{\Gamma^{\rm sf}_{\gamma}(+)\, +\, \Gamma^{\rm sf}_{\gamma}(-)}
\qquad
\tau_{p}^{-1} = \Gamma^{\rm sf}_{\gamma}(+)\, +\, \Gamma^{\rm sf}_{\gamma}(-)
\,.
\eeq

Sokolov and Ternov calculate $\gsf{\gamma}(u)$ in a constant and uniform background magnetic field, $H$,~\cite{Sokolov:1963zn} (see also~\cite{Zhukovskii1971}). In natural units,\footnote{
Here we apply a the conventional particle-physics system of ``natural units'', with $\hbar=c=1$.
} their result for electrons reads
\beq
\label{eq:gsf_ST}
\gsf{\gamma}(u) \Bigg|_{\rm{ST}} = e^2\frac{E_i}{m \rho}\,\frac{5\sqrt3}{144 \pi}
\left(1+ \frac{8}{5\sqrt3}u\right)\xi_0^2\, ,
\eeq
where $e$ is the QED coupling, $E_i$ is the beam energy, $m$ is the electron mass, and $\rho$ is the radius of a circular and planar storage-ring, $e H = \beta E_i / \rho \approx E_i / \rho$. The parameter,
\beq
\label{eq:xi0_def}
\xi_0  \equiv 
\frac{\tfrac{3}{2} e H E_i}{m^3} 
\, ,
\eeq
characterizes the importance of quantum effects in synchrotron radiation.
In storage-rings, $\xi_0 \ll 1$ as terrestrial magnetic fields are small (in natural units).\footnote{
This is not the case for example in pulsars~\cite{Skobelev:2000az,Borisov:1994wg} and other astrophysical systems.
}
Using \eqref{pol_obs_def} one finds~\cite{Sokolov:1963zn,Sokolov:1986nk,Jackson:1975qi,Chao:1981it,Chao:384825, Mane:2005xh,Barber:2006}
\beqa
\label{eq:pol_tau_ST}
|{\cal P}_{eq}|\bigg|_{ST} &\equiv&
|{\cal P}_{ST}| = \frac{8}{5\sqrt3} \approx 92.4\%
\nn\\
\tau_{p}\bigg|_{ST} &\equiv& 
\tau_{ST} =
\left(
e^2\frac{E_i}{m \rho} \,\frac{5\sqrt3}{72 \pi}
\xi_0^2
\right)^{-1}
\,.
\eeqa
The spins of an electron beam tend to align anti-parallel to the magnetic field. Positron beams do the reverse, a result obtained by taking $u\to-u$ in \eqref{gsf_ST}.

\section{Modifications of the Sokolov-Ternov effect in Real storage-rings}
\label{sec:ST_real}

\subsection{Effective Radius}
\label{sec:eff_rad}
Realistic storage-ring designs are not circles with a uniform and constant magnetic field.
Typical designs have both straight and curved sections, respectively known as {\it insertions} and {\it arcs}, with bending magnetic fields predominantly in the latter.
This modifies the circular ring radius, $\rho$, in \eqref{pol_tau_ST} to an effective radius~\cite{Jackson:1975qi} given by
\beq
\label{eq:rho_eff}
\rho_{\rm eff} = \left( \frac{\oint |\rho(s)|^{-3} ds}{\oint ds} \right)^{-1/3} = \left(\rho^2 R\right)^{-1/3}
\eeq
where $\rho(s)$ is the bending radius as a function of the ring contour parameterization, $s$.
The last equality is for a ring consisting of insertions ($\rho\to\infty$), and arcs of equal and constant bending radius $\rho$, and circumference given by $2\pi R$, where $R$ is the effective geometric radius. 
Additional modifications occur for the equilibrium polarization which also take into account the direction of the magnetic field in each arc~\cite{BaierKatkovStrakhovenko}.

\subsection{Real Beam Optics}

Machine deliverables require an intricate system of non-uniform and time-dependent electromagnetic field configurations resulting in the so called ``beam-optics''~\cite{Chao:384825, Wiedemann2015}. Some facilities also include a series of insertion devices like wigglers and undulators which increase synchrotron-radiation production rates or focus them to specific frequencies~\cite{Bilderback_2005,COUPRIE20143}.
The emission of synchrotron radiation in such environments induces a natural depolarization effect known as ``spin-diffusion''~\cite{Chao:1981it,Mane:2005xh, Barber:1999xm, Chao:384825, Barber:2006}. The ST-polarization and spin-diffusion-depolarization are competing effects, which typically result in a lower asymptotic polarization, and shorter polarization time.
A detailed discussion is given in the supplemental materials. 
For the ``quiet'' planar rings we consider in this work, and given a set of machine running conditions which are away from spin-orbit resonances~\cite{Froissart:1960zz, Baier:1965tr,Barber:2004bi}, the quantities in \eqref{pol_evolution} are modified to~\cite{BaierKatkovStrakhovenko, Derbenev:1973ia}
\beq
\label{eq:depol_obs}
{\cal P}_{eq} = {\cal P}_{ST} \frac{\tau_p}{\tau_{ST}}  + {\cal P}_{kin},
\qquad
\tau_p^{-1} = \tau_d^{-1} + \tau_{ST}^{-1}
\, .
\eeq
The term ${\cal P}_{kin}$ is called the ``kinetic polarization'', and is typically very small for planar rings (see~\cite{Derbenev:1973ia}, and the supplemental material).

In ``quiet'' rings, the characteristic depolarization time, $\tau_d$,  is very long, and the polarization observables come very close to the Sokolov-Ternov prediction.
New physics contributions are therefore constrained so as to not modify these results.
Moreover, if both the polarization and the polarization time are measured, then their ratio, 
\beq
\frac{  {\cal P}_{eq}   } { \tau_p }  =  \frac{  {\cal P}_{ST}   } { \tau_{ST} }\,, 
\eeq
can be used to constrain various new physics contributions.
In practice, the time measurement resolution is superior to that of the asymptotic beam polarization, and the $\tau_p$ measurements put stronger constraints.


\section{New Physics Effects on Polarization Observables}
\label{sec:new_physics}

The polarization time in \eqref{depol_obs} would be modified by the synchrotron-like spin-flipping emission of an ultralight DM particle, $X$, 
\beq
\label{eq:tau_with_NP}
\tau_p^{-1} = \tau_d^{-1} + \tau_{ST}^{-1} + \tau_{X}^{-1}\,.
\eeq
In order to calculate $X$'s contribution to the polarization time, $\tau_{X}^{-1}$, we follow in the footsteps of Sokolov \& Ternov's original calculation using the ``method of exact solutions'' approach. We perform a perturbation theory expansion using exact solutions of the Dirac equation in a uniform, and constant background magnetic field~\cite{Sokolov:1963zn}. The derivation is lengthy, and is therefore kept to the supplemental materials, where we generalize the ST approach in order to take into account massive particles.

Notably, other approaches lead to the same results for the photon case, but are not straightforward to generalize to the massive case. For QED, the use of a {\it semi-classical effective Hamiltonian} approach~\cite{Jackson:1975qi, Derbenev:1973ia} is easier to apply in realistic magnetic field configurations.
The formalism of~\cite{Schwinger:1974sz,Tsai:1974id} is useful for the massive case as well.\footnote{
\cite{Chen:1985sw} applies these methods to calculate Weak gauge-boson synchrotron emission, but incorrectly excludes the Goldstone boson contribution from the loop. Nonetheless, the rates are negligible.
}

\subsection{Simplified Models}
\label{sec:models}
We consider four simplified models in which a new particle $X$ couples to electrons.
In these models $X = \{V^\mu,\,A^\mu,\,S,\,a\}$, i.e. a massive vector, an axial vector, a scalar, and a pseudo-scalar.
The interaction Lagrangians are given by
\beqa
\label{eq:Lag_models}
{\cal L}_{\rm int} \supset 
g_V\bar\psi \gamma^\mu \psi V_\mu, ~
g_A\bar\psi \gamma^\mu \gamma^5 \psi A_\mu, ~
g_S\bar\psi  \psi S, ~
i\,g_a\bar\psi \gamma^5 \psi a~~~
\eeqa
where canonically normalized kinetic terms, and mass terms, are implicit.

\subsection{$X$-Induced Spin-Flip Rates}
\label{sec:spin_flip_rates}

Following the notations of \secref{ST_basics}, we calculate the transition-rate for an initial state electron to emit a $X$-particle, such that the final state electron spin has flipped,
\beq
\Gamma^{\rm sf}_{X} \equiv
\Gamma_H(e^-(u) \to e^-(-u) X) \, .
\eeq
For a massive $X$, these rates are attenuated when the value of the ``phase-space'' parameter, $\frac{m_X}{m \,\xi_0}$ exceeds unity. The full results are given as differential transition rates in the supplemental material, and have to be numerically integrated. While we employ them in this analysis, it is instructive to explore the ``massless'' limit, $\frac{m_X}{m \,\xi_0}\to0$, for which $\Gamma^{\rm sf}_{X}$ is maximized. In this limit, the calculation can be performed analytically, and presented as an expansion in $\xi_0$. We find,
\beqa
\gsf{V_T} &=& g_V^2\,\frac{5\sqrt3}{144 \pi}
\frac{E_i}{m \rho}
\left(1+ \frac{8}{5\sqrt3}u\right)\xi_0^2
\\
\gsf{V_L} &=& g_V^2\,\frac{5\sqrt3}{144 \pi}
\frac{E_i}{m \rho}
\left(\frac{m^2m_V^2}{108\,E_i^4}\right)
\left(1+ \frac{8}{5\sqrt3}u\right)\xi_0^2
\\
\label{eq:gsf_AT}
\gsf{A_T} &=& g_A^2\,\frac{1}{144 \pi}\frac{E_i}{m \rho}
\left(
36 (\sqrt3 + u)
\right)
\\
\label{eq:gsf_AL}
\gsf{A_L} &=& 
g_A^2\,\frac{\sqrt3}{162 \pi}\frac{E_i}{m \rho}
\left(85+48\sqrt3\,u\right)\frac{m^2}{4m_A^2}\xi_0^2
\\
\gsf{S} &=& g_S^2\,\frac{5\sqrt3}{162 \pi}\frac{E_i}{m \rho}\xi_0^2
\\
\gsf{a} &=& g_a^2\,\frac{\sqrt3}{162 \pi}\frac{E_i}{m \rho}
\left(85+48\sqrt3\,u\right)\xi_0^2
\eeqa
where for the massive vector, and axial-vector cases we distinguish between the transverse, and longitudinal modes, denoting them by $_T$, and $_L$ respectively.

A few points are noteworthy here.
First, our result for $\gsf{V_T}$ agrees with the ST one, see \eqref{gsf_ST}.
Second, the results for the longitudinal modes behave as expected.  
In the axial-vector case, $\gsf{A_L} = \left(\tfrac{m^2}{4m_A^2}\right) \gsf{a}$, which is consistent with the Goldstone Boson Equivalence theorem. In the vector case, $\gsf{V_L}$ has the correct decoupling properties~\cite{An:2013yfc}, but otherwise preserves the spin-flip structure of the transverse mode.
Third, while scalars in a magnetic field can flip the spin, the flip rate is insensitive to the spin-state, and so polarization does not build up overall.\footnote{In fact, for a scalar, all $u\to u'$ transitions have the same rate.} Most importantly, with the exception of $\gsf{A_T}$, all the spin-flip rates start at the $\xi_0^2$ order ! 
We stress that the previous statement also includes $\gsf{A_L}$ which is enhanced~\cite{Dror:2018wfl}, but subject to the unitarity constraint~\cite{Kahlhoefer:2015bea, Dror:2020fbh}
\beq
\label{eq:unitarity_constraint}
g_A \frac{m}{m_A} \lesssim 
\sqrt{\tfrac{\pi}{2}} \,.
\eeq 
The lack of $\xi_0$ (magnetic field) suppression in the transverse axial-vector spin-flip rate arises due to its spin-parity quantum numbers. It can be readily understood by noting that in the non-relativistic limit, an axial-vector couples directly to the electron spin.

The experimental precision for polarization measurements is at the sub-to few-percent level~\cite{Aulenbacher:2018weg}. If $\gsf{X}$ is at the same order in $\xi_0$ as the SM contribution, then the corresponding $X$ sensitivity would not be competitive with existing searches.
We therefore focus on the {\it axial-vector} case~\cite{Kahn:2016vjr}.

\section{Available Experimental Results}
\label{sec:available_measurements}

Beam polarization measurements have been performed in a plethora of machines with energies in the range $\sim 0.5~\gev~-~45~\gev$~\cite{Shatunov:2006ng, Mane:2005xh, Gianfelice-Wendt:2018mcx}. 
We focus on ``quiet'' low-energy storage-rings, such as third-generation synchrotron light-sources. These are particularly interesting because the machine running conditions can be tuned to have negligible depolarization effects, such that the polarization-time nearly saturates the Sokolov-Ternov prediction.
\tableref{exp_data} presents the polarization measurements used in this work. The last column, gives a conservative $2\sigma$ estimate for the sum of depolarization effects: 
\beq
\label{eq:gamma_max}
\Gamma^{\rm max}_X = \left(\tau_{p} - 2\Delta\tau_p\right)^{-1} - \tau_{ST}^{-1}\,,
\eeq
which is used in \secref{limits} for limit setting.
\begin{table*}
\center
\begin{tabular}{|c|c|c|c|c|||c|c||c|}
\hline
\hline
Exp & $E_i$ [GeV] & $\rho$~[m] & $2\pi R$ [m] & $\xi_0$ & $\tau_{ST}$ & $\tau_p$ & 
$ \Gamma^{\rm max}_X~[\gev]$ \\
\hline
\hline
SLS~\cite{Leemann:2002br} & 2.4 & 11.48 & 288  & $2.22\times10^{-6}$ & $1873~\rm{sec}$ & $1837\pm1~\rm{sec}$ & $7.478\times 10^{-30}$\\
\hline
SPEAR3~\cite{Wootton:2013lma}& 3.0 & 8.144 & 234.144  & $2.44\times10^{-6}$ & $1003~\rm{sec}$ & $840\pm17~\rm{sec}$& $1.605\times 10^{-28}$	 \\
\hline
\hline
\end{tabular}
\caption{Summary of experimental data on beam polarization measurements used in this work.}
\label{table:exp_data}
\end{table*}
%
While other measurements exist (for example~\cite{Steier:2000ui, Birkel:2004jk, Sun:2010zzi, Kuske:2010zza, Zhang:2013zw, Carmignani:2015jwi, Vitoratou:2019igw}), we prioritize those which have the highest precision, and provide standard errors on fit parameters.

The SLS~\cite{Leemann:2002br}, and SPEAR3~\cite{Wootton:2013lma} measurements employ the resonant spin-depolarization technique~\cite{Baier_1972, Derbenev:1980gp}, a standard beam polarimetry technique which is often used for accurate calibration of the beam energy.
In resonant spin-depolarization, a radio-frequency magnetic field is turned on in the plane of the ring. The frequency of the field is set to a spin resonance, an integer times $a_e\gamma$~\cite{Baier:1965tr}, which causes the beam to depolarize. When the field is turned-off, the beam begins to polarize again. 
Due to the Touschek effect~\cite{Bernardini:1997sc,Streun2001}\footnote{
Intra-beam collisions (M{\o}ller scattering) of oppositely polarized electrons have a higher cross-section and therefore lead to a higher beam loss rate (particles emitted out of the beam).
}
the beam lifetime is correlated with its polarization, and $\tau_p$ can be measured by monitoring beam losses as those decrease after the beam has been depolarized.
Note that this technique only measures $\tau_p$, while $P_{eq}$ is deduced assuming the SM, i.e. assuming the ST-effect (though see~\cite{Lee:2005xx}).
In addition, the dependence of the Touschek scattering on the beam polarization is not modified by an additional weakly coupled axial-vector force that mediates the reaction.

Other polarimetry techniques exist with potentially higher precision~\cite{Aulenbacher:2018weg}. We mention in passing the Compton backscattering based polarimeters~\cite{Gustavson:1979fa}\footnote{
Coherent laser scattering off the beam exhibits polarization dependent energy and detector-plane hit distributions. By monitoring this over time, the polarization observables, \eqref{depol_obs}, can be measured.
}, including the Fabry-Perot cavity used in the final days of HERA~\cite{Sobloher:2012py}.
These two techniques measure the polarization directly, rather than just its relative build-up over time.
We refrain from using these measurements in this work. To the best of our knowledge, the inverse-Compton-based methods were not used in sufficiently ``quiet'' machines, while the HERA measurements were done at relatively high-energy, with large depolarization effects.


\section{Experimental Reach}
\label{sec:limits}

\begin{figure}[h!]
\centering
\includegraphics[width=0.5\textwidth]{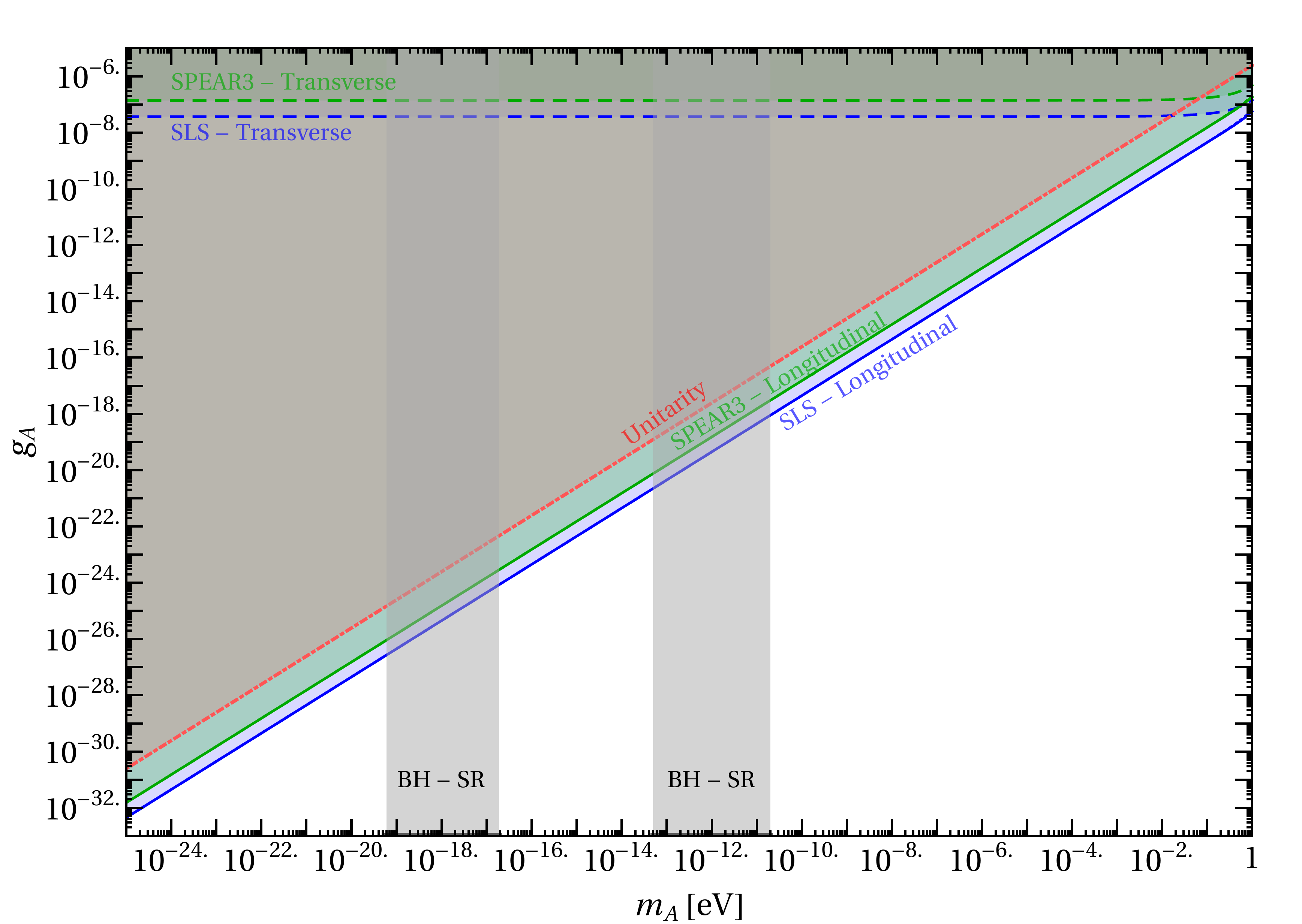}
\caption{
Limits on light axial-vectors coupled to electrons in the $\{g_A,~m_A\}$ plane.
Shaded regions are excluded. 
The triangular exclusion (constrained by the red doted-dahsed line) corresponds to the unitarity constraint of \eqref{unitarity_constraint}. The two gray bands correspond to limits from
black hole super-radiance (BH-SR)~\cite{Baryakhtar:2017ngi}.
Other possible limits are referred to in the text.
The SLS and SPEAR3 constraints are given in blue and green respectively. Dashed lines correspond to the transverse contributions, while the continues ones are the sum of the transverse and longitudinal ones.
}
\label{fig:limits_axial_vector}
\end{figure}

Limits on new axial-vectors coupled to electrons are set by requiring that their contribution to the spin-flip rate does not exceed the sum of all depolarizing contributions which is estimated by \eqref{gamma_max},
\beq
\gsf{X} < \Gamma^{\rm max}_X
\eeq
The discussion of \secref{eff_rad} is taken into account by applying a global rescaling factor, $\rho/R$ to the $\gsf{X}$ full results for (see also \eqsref{gsf_AT}{gsf_AL}).
\figref{limits_axial_vector} shows the resulting limits in the $\{g_A,~m_A\}$ plane.
Notably, these strong limits represent a conservative estimate as they assume negligible machine induced depolarization effects. 
We point out that $\tau_d$ could be estimated using dedicated algorithms and tracking codes~\cite{Chao:1980fz, Yokoya:1992ti,  Eidelman:1993qq,  Heinemann:1996ig,  Yokoya:1999ip,  Luccio:1999wd,  Mane:2003uz,  Barber:2004bi, Heinemann:2015wwa,  Abell:2015gfa,Duan:2015wja} to a~$10\%$ accuracy~\cite{Barber_priv} which in turn, would lead to stronger limits.

While the SLS and SPEAR3 facilities have $\xi_0\sim2\times10^{-6}$, the SLS limit is stronger due to a quieter environment, and a more precise measurement (\tableref{exp_data}). 
In addition, both limits are stronger than the unitarity requirement, \eqref{unitarity_constraint}. 
The reach is dominated by the longitudinal-mode contribution up to the $\xi_0\, m \approx 1~\ev$ scale where phase-space effects come into play, and the transverse mode contributions (dashed) become comparable.

The limits obtained here are model independent, and require only that a spin-1 particle couples to $e^\pm$ through an axial-vector coupling. In concrete models with a spin-1 particle, a vector coupling imposes additional constraints~\cite{Goodsell:2009xc, Knapen:2017xzo,An:2013yfc, Redondo:2015iea}. Some of these constraints may also apply directly to the axial-vector couplings, however, such an analysis is beyond the scope of this work.
Interestingly, $5^{\rm th}$-force experiments and equivalence-principle tests~\cite{Wise:2018rnb} are not sensitive to axial-couplings because the low-energy potential induced by an axial-vector couples spins, and therefore averages to zero. Notwithstanding, black-hole superradiance signatures~\cite{Baryakhtar:2017ngi} may limit the axial-vector coupling directly. 
Moreover, if the longitudinal mode of the axial-vector is interpreted as an axion-like particle, then additional bounds may exist from the corresponding searches~\cite{Raffelt:1996wa, Ringwald:2014vqa, Battaglieri:2017aum}.
Finally, this analysis is complementary to~\cite{Arcadi:2019uif}, which constrains $Z'$-models down to~${\cal O}(10~\eV)$ using atomic parity violation experiments.
\\

\section{Conclusions and Outlook}
\label{sec:conclusions}

The Sokolov-Ternov effect, the radiative self-polarization of $e^\pm$-beams, is a well tested phenomenon in storage-rings.\footnote{See~\cite{Graham:2020kai, Janish:2020knz} for recent storage-ring based direct-detection experiments} Synchrotron-like emission of new light particles off $e^\pm$ can modify polarization observables, and is therefore testable using polarization measurement.
In ``quiet'' storage-rings, depolarization effects are small, and polarization measurements are sufficiently accurate to constitute as precision tests of the effect.

This work explores the sensitivity of these measurements to new physics.
We calculate the spin-flip transition-rates following the Sokolov-Ternov approach, by generalizing the ``method of exact solutions'', and the associated approximations to include massive particles.
We find that the characteristic polarization time is particularly sensitive to spin-flips by axial-vectors which are not suppressed by the smallness of laboratory magnetic fields, in contrast to particles with other spin-parity assignments.

Limits are set by requiring that the axial-vector contribution to the polarization time does not exceed that of the measured sum of depolarization effects.
This conservative approach results in extremely strong limits on light axial-vectors coupled to electrons. Nonetheless, there is potential for improvement by applying dedicated storage-ring tracking tools which can reliably estimate the depolarization effects, namely, $\tau_d$.

\begin{acknowledgments}
I am grateful to
Tom Banks,
Matt Buckley,
Ranny Budnik,
Jeff Dror,
Marat Freytsis,
Yuri Gershtein,
Enrique Kajomovitz,
Simon Knapen,
Zohar Komargodski,
Stefan Schmitt,
Yael Shadmi,
David Shih,
Yotam Soreq,
and
Scott Thomas,
for the many fruitful discussions which have benefited this work.
I am particularly grateful to Desmond Barber, for a careful read of the manuscript, and for many fruitful discussions.
Most of all, I am grateful to my family for their support in these challenging and uncertain times.
This work was supported in part by DOE grant DE-SC0010008.
\end{acknowledgments}

\twocolumngrid
\vspace{-8pt}
\section*{References}
\vspace{-10pt}
\def\bibsection{}
\bibliographystyle{utphys}
\bibliography{polarization}

\clearpage
\newpage
\maketitle
\onecolumngrid

\begin{center}
\textbf{\large } \\
\vspace{0.05in}
\textit{\large Supplemental Material}\\
\vspace{0.05in}
{Iftah Galon}
\end{center}

\onecolumngrid
\setcounter{equation}{0}
\setcounter{figure}{0}
\setcounter{table}{0}
\setcounter{section}{0}
\setcounter{page}{1}
\makeatletter
\renewcommand{\theequation}{S\arabic{equation}}
\renewcommand{\thefigure}{S\arabic{figure}}
\renewcommand{\thetable}{S\arabic{table}}

\section{Polarization and Spin Dynamics in Storage Rings}
\label{sec:depol_effects}

Spin dynamics and polarization in accelerator physics is an intricate topic.
This section summarizes key points which are used in this work.
The interested reader is referred to reviews by~\cite{Mane:2005xh,Hoffstaetter:2006zu}, and to the useful lecture notes by~\cite{Barber:2006}.

In accelerator physics, particle dynamics is governed by an accelerator Hamiltonian.
If the electromagnetic fields are sufficiently spatially homogeneous, Stern-Gerlach forces can be neglected, which implies that the spin does not back-react on the orbital dynamics. In these cases, the general form of the quantum Hamiltonian can be written as~\cite{Mane:2005xh,Hoffstaetter:2006zu}
\beq
\hat{\cal H} =\hat{ {\cal H}}_{orb} + \hat {\vec W}\cdot \hat{ \vec S }
\eeq
where ${\cal H}_{orb}$ is the Hamiltonian for the orbital dynamics, and $\vec S$ is the quantum spin operator, defined in the rest frame of the particle (hats denote operators).
In the semi-classical approximation approach the Ehrenfest theorem is applied (see also~\cite{Barber:2006}) to obtain an equation of motion for the {\it classical spin-vector}, $\vec s = \braket{\hat{\vec S}}$,
\beq
\label{eq:spin_evol}
\frac{d\vec s}{dt} = \vec \Omega \times \vec s
\eeq
where a ``classical'' trajectory is assumed, i.e. $\vec \Omega = \braket{ \,\hat{\vec W} (   \hat{\vec q},\,\hat{\vec p} )\, }
\approx
\vec W \left( \, \braket{\vec q},\, \braket{\vec p} \, \right)\, .
$
This is a good approximation for most low-energy storage rings as the characteristic orbital times are much shorter than those for the spin polarization, see~\cite{Montague:1983yi, Barber:2006}.
For spin-physics, the orbital phase space can be considered {\it ergodic}.
In the case of single particle dynamics, in constant and uniform electromagnetic fields, the $\Omega$ in \eqref{spin_evol} is given by the Thomas-Bargmann-Michel-Telegdi equation~\cite{Thomas:1927, Bargmann:1959gz},
\beq
\Omega_{TBMT} = -\frac{e}{m}\left[
\left(a+\frac1\gamma\right)\vec B
-\frac{a\gamma}{\gamma+1}(\beta\cdot\vec B)\vec \beta
-\left(a+\frac1{\gamma+1}\right)\vec\beta\times\vec E
\right]
\eeq

The approximation breaks down when radiation is involved because the time-scale for radiation emission is much shorter than any orbital time-scale. To take radiation into account one uses the {\it semi-classical effective Hamiltonian for QED} approach~\cite{Jackson:1975qi, Derbenev:1973ia}. The effective single-particle Hamiltonian is obtained by extending the background field interactions to the radiation fields,\footnote{
Alternatively, by a Foldy-Wouthuysen transformation~\cite{Barber:2006}
}  i.e.
\beq
\label{eq:semiclassical}
\hat {\cal H}_{\text{{\it semi-classical} QED}}
=
\left(
(\hat{\vec p} - e \hat{\vec A})^2 + m^2
\right)^{1/2}
+e \hat V + \hat{\vec \Omega}\cdot \hat {\vec s}\, .
\eeq
Here the electromagnetic field dependent terms are denoted as (hatted) operators because they involve both the background (classical) field, and the quantized radiation field (operator),
\beq
\vec A = \vec A^{\rm{bkg}} + \vec A^{\rm{rad}},
\qquad
V = V^{\rm{bkg}} + V^{\rm{rad}},
\qquad
\vec \Omega = \vec \Omega^{\rm{bkg}} + \vec \Omega^{\rm{rad}} \,
\eeq
and the Hamiltonian can be expanded in powers of the radiation field.

The asymmetry in the spin-flip transition rates contributes to polarization build up. In equilibrium, the beam polarization, generally written as $\vec {\cal P} = \frac1{N}\sum_{\rm i=1}^N\braket{\vec s_i}$, can be written as the average~\cite{BaierKatkovStrakhovenko, Derbenev:1972mk, Baier_1972, Derbenev:1973ia, Mane:2005xh}
\beq
\label{eq:Peq}
|\vec {\cal P}_{eq}| = |\braket{\braket{\vec s \cdot \hat n}\hat n}|
=
\frac{\braket{\Gamma_+ - \Gamma_-}}{\braket{\Gamma_+ + \Gamma_-}}
\eeq
The vector $\hat n = \hat n(z,\,\theta)$, known as the {\it invariant spin-field}~\cite{Hoffstaetter:2006zu}, is a field of spin-quantization axes on the generalized phase-space point, $z=z(\vec q,\,\vec p)$, and generalized azimuthal coordinate along the ring.
The invariant spin-field is a solution of \eqref{spin_evol}, required to be one-turn periodic. Namely, after one turn, i.e. $\theta\to\theta+2\pi$, a particle phase-space location evolves from its initial phase-space point, $z_i$, to $z_f$, a new phase-space location. The requirement on $\hat n$ is expressed as
\beq
{\cal M} \hat n(z,\, \theta) = \hat n({\cal M}z,\,\theta)
\eeq
where ${\cal M}$ performs the evolution of the system in the spin, and orbital spaces, and is known as the {\it one-turn spin-orbital map}~\cite{Mane:2005xh,Hoffstaetter:2006zu}.
In \eqref{Peq} the spin projection onto the quantization-axis is averaged over the local phase-space volume element at $z$ (inner average), and the outer average is over the entire orbital phase-space. This is then expressed as a balance of spin-flip transition rates. In the second equality, the averages are of spin-flip transition rates over the entire orbital phase-space.

Derbenev and Kondratenko~\cite{Derbenev:1972mk,Derbenev:1973ia} show how to perform the averages in \eqref{Peq}, taking into account general electromagnetic field configurations using the invariant spin-field,
\beqa
\label{eq:DK_formulas_P}
{\cal P}_{DK} &=& {\cal P}_{ST}
\frac{
\braket{
\oint ds
\tfrac{1}{|\rho|^3}\hat b\cdot\left(\hat n - \tfrac{\partial \hat n}{\partial \delta}\right)
}
}
{
\braket{
\oint ds
\tfrac{1}{|\rho|^3}
\left(
1 - \tfrac29(\hat n\cdot\hat v)^2 
+ \tfrac{11}{18}
\left|\tfrac{\partial \hat n}{\partial \delta}\right|^2
\right)
}
}
\\
\label{eq:DK_formulas_tau}
\tau_{DK}^{-1} &=&
\tau_{ST}^{-1}
\braket{
\oint ds
\tfrac{1}{|\rho|^3}
\left(
1 - \tfrac29(\hat n\cdot\hat v)^2 
+ \tfrac{11}{18}
\left|\tfrac{\partial \hat n}{\partial \delta}\right|^2
\right)
}
\eeqa
where $\hat b = \frac{\hat v \times \dot{\hat{v}}    }{ | \hat v \times \dot{\hat{v}}    |}$, $\hat v$ is a unit vector in the direction of the local velocity, $\dot{\hat{v}} $ is its time-derivative, and $s$ is a length element along the storage ring. The average is performed on the orbital phase-space ($z$). The $\tfrac{\partial}{\partial\delta}$ derivatives are with respect to $\delta = \Delta p/p$, the relative momentum offset, one of the orbital-phase space coordinates.

In most ``quiet'' planar rings, the product $\hat n\cdot \hat v$ is negligible.
Neglecting $\hat n\cdot \hat v$ terms in \eqsref{DK_formulas_P}{DK_formulas_tau}, and defining
\beq
{\cal P}_{kin} = -{\cal P}_{ST}\frac{\tau_{DK}}{\tau_{ST}}\braket{
\oint ds
\tfrac{1}{|\rho|^3}
 \hat b\cdot \tfrac{\partial \hat n}{\partial \delta}
}
\,, \qquad
\tau_d^{-1} = 
\tau_{ST}^{-1}
\braket{
\oint ds
\tfrac{1}{|\rho|^3}
 \tfrac{11}{18}
\left|\tfrac{\partial \hat n}{\partial \delta}\right|^2
}
\,,
\eeq
one obtains \eqref{depol_obs}. 
In ``quiet'', planar rings, $\hat n$ is close to vertical, and $\partial \hat n/\partial \delta$ is small.
As a result, the kinetic polarization is small !

In a photon emission process, the initial and final charged particle states have different energies ($E_i$, and $E_f$ respectively).
As a result, the spin quantization axes of the initial state, $\ket{n(E_i)}$, and the final state, $\ket{n(E_f)}$ are misaligned, and the initial and final fermion states are not orthogonal.
With this observation, Derbenev and Kondratenko conclude that the spin-flip transition rate draws from two contributions in the semi-classical QED effective theory~\eqref{semiclassical}.
The first is the spin-dependent term, $\vec\Omega_{rad}\cdot\vec s$, and the second is from the usual covariant derivative term. The latter, is typically diagonal in the spin sub-space, but due to the misalignment of the spin-quantization axes it can non-trivially couple the initial and final spin-states. While this effect is sub-leading with respect to the spin-conserving photon emission, it is of the same order of magnitude as that coming from the $\vec\Omega_{rad}\cdot\vec s$ term.

For new physics with a direct spin-dependent coupling at leading order, as in the case of an axial-vector, a similar effect would be sub-leading, and so we ignore its possible contributions.


\section{Notations and Conventions}
\label{sec:notations}
Greek letters are used to denote space-time indices running over $0,1,2,3$.
Roman letters are used to denote space indices running over $1,2,3$.
The space-time Minkowski metric is chosen with mostly negative signature
\beq
g_{\mu\nu} = \rm{diag}\left[1,~-1,~-1,~-1\right]
\eeq
For the Dirac $\gamma$-matrices we use the Chiral representation:
\beq
\gamma^{\mu} = 
\begin{pmatrix}
0 & \sigma^\mu \\
\bar\sigma^\mu & 0
\end{pmatrix}
\eeq
where $\sigma^\mu = \left( 1_{2X2},\sigma^i\right)$, and $\bar\sigma^\mu = \left( 1_{2X2},-\sigma^i\right)$ are four-vectors of $2X2$ matrices, $\sigma^{i}$ are the Pauli matrices which are given by,
\beq
\sigma^1
=
\begin{pmatrix}
0 & 1 \\
1 & 0
\end{pmatrix}
,\quad
\sigma^2
=
\begin{pmatrix}
0 & -i \\
i & 0
\end{pmatrix}
,\quad
\sigma^3
=
\begin{pmatrix}
1 & 0\\
0 & -1
\end{pmatrix}\, .
\eeq
Explicitly,
\beq
\gamma^{0} = 
\begin{pmatrix}
0 & 0 & 1 & 0 \\
 0 & 0 & 0 & 1 \\
 1 & 0 & 0 & 0 \\
 0 & 1 & 0 & 0
\end{pmatrix},
\quad
\gamma^{1} = 
\begin{pmatrix}
0 & 0 & 0 & 1 \\
 0 & 0 & 1 & 0 \\
 0 & -1 & 0 & 0 \\
 -1 & 0 & 0 & 0 
\end{pmatrix},
\quad
\gamma^{2} = 
\begin{pmatrix}
0 & 0 & 0 & -i \\
 0 & 0 & i & 0 \\
 0 & i & 0 & 0 \\
 -i & 0 & 0 & 0 
\end{pmatrix},
\quad
\gamma^{3} = 
\begin{pmatrix}
 0 & 0 & 1 & 0 \\
 0 & 0 & 0 & -1 \\
 -1 & 0 & 0 & 0 \\
 0 & 1 & 0 & 0
\end{pmatrix},
\quad
\gamma^{5} = 
\begin{pmatrix}
 -1 & 0 & 0 & 0 \\
 0 & -1 & 0 & 0 \\
 0 & 0 & 1 & 0 \\
 0 & 0 & 0 & 1
\end{pmatrix}
\eeq
The four-vector derivative operator is given by
\beq
\partial_\mu = \frac{\partial}{\partial x^\mu} = \left(\frac{\partial}{\partial x^0},~\vec\nabla\right)
\eeq
the kinetic four-momentum representation is given by $p^\mu = i \partial^\mu$.
The spin-matrices are given by
\beq
\Sigma^k = 
\begin{pmatrix}
\sigma^k & 0 \\
0 & \sigma^k 
\end{pmatrix}
\eeq

\section{Dirac Equation Solution in a Background Constant \& Uniform Magnetic Field, in Cylindrical coordinates}
\label{sec:Dirac_eq_sols_short}

We consider Quantum Electrodynamics with an external background magnetic field as the unperturbed theory. For this theory the Lagrangian is given by
\beq
{\cal L} = \bar\psi\left(
i\slashed{\partial} - e Q\slashed{A}^{\rm{bkg}} - m
\right)\psi
-\tfrac14F^{\mu\nu}F_{\mu\nu}
\eeq
In this theory, momentum is conserved only along the $z$-axis.
The Dirac field can be quantized using the wave function solutions to the E.O.M's with positive-energy (particle) and negative-energy (anti-particle) states.
These wave function solutions are given by
\beq
\label{eq:gen_sol}
\psi_{\epsilon_p}(r,\,\phi,\,z,t) = \exp\left(
- i\,\epsilon_p E\, t
\right)
\frac{\exp\left(
i\, \epsilon_p \,k_3\, z
\right)}{\sqrt L}
\frac{\exp\left( i(\ell - \tfrac12)\phi\right)}{\sqrt{2\pi}}
\exp\left(-i\tfrac12\phi\Sigma_3\right)
\sqrt{2\gamma}
\begin{pmatrix}
c_1\,\ILag{n-1}{s}{\rho} \\
c_2\,\ILag{n}{s}{\rho} \\
c_3\,\ILag{n-1}{s}{\rho} \\
c_4\,\ILag{n}{s}{\rho}
\end{pmatrix}
\eeq
where we have defined,
\beq
\label{eq:gamma_rho}
\gamma = \tfrac{1}{2}e Q H,
\qquad
\rho = \gamma r^2
\eeq
and where $\epsilon_p=\pm$ for particle/anti-particle, and $Q$ is the corresponding charge in units of $(-e)$.
For the z-axis we take a one-dimensional box with edges at $\pm L$, which can later be taken to infinity.
The Lagurre gaussian functions are\footnote{
Note that these are properly normalized, i.e. $\int_0^\infty (\ILag{n}{s}{\rho})^2 = 1$
}
\beq
\ILag{n}{s}{\rho} = \sqrt\frac{1}{n!s!} e^{-\rho/2}\,\rho^{\frac{n-s}{2}}\, Q^{n-s}_s(\rho)
\eeq
where $n-s = \ell$ is a positive integer called the {\bf Landau-level}, $\ell,\,s$ are integers, and the $Q$ functions are the Lagurre polynomials.
The $c_i$'s are chosen such that $\sum_i |c_i|^2=1$, and $\psi$ is chosen as an eigenstate, such that 
\beq
\frac{m}{\sqrt{E^2-k_3^2}}\,\mu_3 \psi = u\psi, \qquad  u=\pm1
\eeq
where $\mu_3$ is the magnetic polarization in the $3$-direction which is defined by
\beq
\mu_3 = \Sigma^3 -i \epsilon^{3,j,k}\gamma^j \frac{P^k}{m}
\eeq
here $P = p + e A^{bkg}$ is the canonical momentum.
\beqa
\label{eq:ci_s}
\begin{matrix}
c_1^{\epsilon_p} = \tfrac12 \left(1-\frac{k_3}{E}\right)^{1/2} 
\left(1 + \frac{m\,u}{\sqrt{E^2-k_3^2}}\right)^{1/2}
& ~~&
c_3^{\epsilon_p} = +\tfrac{1}2 u\,\epsilon_p\left(1+\frac{k_3}{E}\right)^{1/2} 
\left(1 + \frac{m\,u}{\sqrt{E^2-k_3^2}}\right)^{1/2}
\\
c_2^{\epsilon_p} = -\tfrac{i}{2}\,\epsilon_p \left(1+\frac{k_3}{E}\right)^{1/2} 
\left(1 - \frac{m\,u}{\sqrt{E^2-k_3^2}}\right)^{1/2}
& ~~&
c_4^{\epsilon_p} = +\tfrac{i\,}2 u\left(1-\frac{k_3}{E}\right)^{1/2} 
\left(1 - \frac{m\,u}{\sqrt{E^2-k_3^2}}\right)^{1/2}
\end{matrix}
\nn\\
\eeqa
The energy eigenvalues are given by
\beq
E \equiv E_{n,k_3} = \left(m^2 + k_3^2 + 4\gamma n\right)^{1/2} \,
\eeq

\section{Field Quantization}
\label{sec:field_qunat}

The Dirac field is quantized such that,
\beq
\label{eq:psi_field}
\psi(t,r,\phi,z)=
\int \frac{dk_3}{2\pi\sqrt{2E_{n,k_3}}}
\sum_{n,s,u=\pm1}
\frac{e^{   i\ell\,\phi   }}{\sqrt{2\pi}}
\left(
e^{   -i\left(E_{n,k_3}t - k_3z\right)   }
U(n,s,k_3,u) a_{n,s,k_3,u}
+
e^{   +i\left(E_{n,k_3}t - k_3z\right)   }
V(n,s,k_3,u) b^\dagger_{n,s,k_3,u}
\right)
\eeq
where
$U$ and $V$ are the {\it radial} four component spinor wave-functions from \eqref{gen_sol},
\beq
\begin{cases}
U \\ V
\end{cases}
=
\sqrt{2E_{n,k_3}}
\sqrt{2\gamma}
e^{-i\tfrac\phi2\left(1_{4\times4} + \Sigma_3 \right)}
\begin{pmatrix}
c_1^\pm\,\ILag{n-1}{s}{\rho} \\
c_2^\pm\,\ILag{n}{s}{\rho} \\
c_3^\pm\,\ILag{n-1}{s}{\rho} \\
c_4^\pm\,\ILag{n}{s}{\rho}
\end{pmatrix}
\,,
\eeq
and $c_i$'s are taken from \eqref{ci_s} with $\epsilon_p=+1,-1$ for $U,V$ respectively.
Note that with this normaliztion
$
\int_0^\infty dr\, r^2 
\{
U^\dagger U,\,
V^\dagger V
\} =2E_{n,k_3}\, .
$
States are normalized such that
\beqa
&&\sqrt{2E_{n,k_3}}\,a^\dagger_{n,s,k_3,u} \ket{0} = \ket{1_{\text{particle}},n,s,k_3,u}
\nn\\
&&\sqrt{2E_{n,k_3}}\, b^\dagger_{n,s,k_3,u} \ket{0} =  \ket{1_{\text{anti-particle}},n,s,k_3,u}
\nn\\
&&
a_{n,s,k_3,u} \ket{0} = b_{n,s,k_3,u} \ket{0} =0
\eeqa
where the corresponding creation/annihilation operators obey the anti-commutation relations
\beq
\{a^\dagger_{n,s,k_3,u}, a_{n',s',k'_3,u'} \} =
\{b^\dagger_{n,s,k_3,u}, b_{n',s',k'_3,u'} \} =
\delta_{nn'}\delta_{ss'}\delta_{k_3k'_3}\delta_{uu'}
\eeq
and with all other anti-commutators vanishing.
These result in
\beq
\label{eq:field_quant}
\{\psi(r,\phi,z),\,\psi^\dagger(r',\phi',z')\} = 1_{4\times4}\frac{1}{r}\delta(r-r')\delta(z-z')\delta(\phi-\phi')
\eeq
The choice of the ``particle physics'' field normalization factor $(2E_{n,k_3})$ is to assure a Lorentz invariant form to boosts along the 3-direction.

For the $X = \{V^\mu,~A^\mu,~S,~a\}$ radiation fields we use free particle solutions. 
For $V^\mu$
\beq
\label{eq:rad_fields}
V_\mu(x,t)=
\int \frac{d^3q}{(2\pi)^3}
\frac{1}{\sqrt{2E_q}}
\sum_{r=0}^3
\left(
a_q^r
\epsilon_\mu^r(q)
e^{-ix\cdot q}
+
a_q^{r\,\dagger}
\epsilon_\mu^{r*}(q)
e^{+ix\cdot q}
\right)
\eeq
and with similar expressions for the other fields.

\section{The $X$-emission rate}

The calculation of the $X$-emission rate involves {\it four} small parameter expansions.
The first is the usual small coupling series expansion of time-dependent perturbation theory.
This is discussed in \appref{X_in_TDPT}. The second, is an expansion of the reduced matrix element in terms of several characteristically small kinematic ratios. This is discussed in \secref{small_param_exp}. The third, is a WKB approximation of the Lagurre-Gaussian functions, $\ILag{n}{n'}{\bar x}$ close to the turning point, where $1 -\bar x/x_0$ is small. This is discussed in \appref{WKB}. In the final evaluation step, it is useful to expand the results as a power series in $\xi_0$ from \eqref{xi0_def}.
The following subsections closely follow \cite{Sokolov:1986nk}, and generalize its results and analyses to the massive $X$ case.

\subsection{$X$-emission in first-order perturbation theory}
\label{sec:X_in_TDPT}

$X$-emission is described in first-order perturbation theory by the interaction Lagrangian
\beq
{\cal L}_{\rm{int}} = \int d^3 x\, \bar\psi (\vec x,t) O_X(\vec x,t) \psi(\vec x,t)
\eeq
where $O_X$ is an operator linear in the $X$-field, with a model dependent spinor representation.
The initial and final states are respectively,
\beqa
\ket{i} &=& \sqrt{2 E_{n,k_3}} \ket{1_{\text{particle}},n,s,k_3,u} \,,
\nn\\
\ket{f} &=& \sqrt{2 E_{n',k'_3}} \ket{1_{\text{particle}},n',s',k'_3,u'}\times \sqrt{2 E_q} \ket{1_X,\vec q} 
\eeqa
where the $X$ 3-momentum, $\vec k$, is chosen in the $YZ$-plane using the azimuthal symmetry of the problem, 
\beq
\vec q = |\vec q|
\begin{pmatrix}
0, & \sin\theta_X, & \cos\theta_X
\end{pmatrix}
 \,.
\eeq

The polarized-$X$-emission rate with initial (final) polarization $u$ ($u'$) is given by
\beq
\label{eq:trans_rate}
\Gamma_X^{u\to u'} = 
\frac{1}{2E_{n,k_3}}
\sum_{n'=0}^n
\sum_{s'=0}^{\infty}
\int \frac{dk'_3}{(2\pi)}
\frac{1}{2E_{n',k'_3}}
\int \frac{d^3q}{(2\pi)^3}
\frac{1}{2E_q}
\left|\int d^3x\,M_{uu'}(r,\phi,z)\right|^2 \,
(2\pi)^2
\,\delta\left(E_i - E_f - E_q\right)
\,\delta\left(k_3 - k'_3 - |\vec q|\cos\theta_X \right)
\eeq 
Here $E_i = E_{n,k_3}$, $E_f = E_{n',k'_3}$, the sums are over $\{n',\,s'\}$, respectively, the final-state fermion Landau-level, and radial quantum numbers, and we have employed the energy and 3-momentum conservation of the problem to obtain the reduced matrix element for $X$-emission from electrons reads
\beqa
M_{uu'}(r,\phi,z) = &&g_X \left(\bar U(n',s',k'_3,u') \tilde O_X \bar U(n',s',k'_3,u')\right)
\exp\left[i\left(
\phi(\ell-\ell')
-r\,|\vec q|\sin\theta_X\sin\phi
\right)
\right]
\nn\\
\eeqa
One performs the $\phi$ and subsequently the $r$ integral using
\beqa
\label{eq:int_dphi}
\frac{1}{2\pi}
&&\int_{0}^{2\pi} d\phi\, \exp\left[i\left(
\phi(\ell-\ell')
-r|\vec q|\sin\theta_X\sin\phi
\right)
\right]
= J_{\ell-\ell'}(|\vec q|\,r\sin\theta_X)
\\
\label{eq:JII}
&&\int_{0}^{\infty} (2\gamma)\,r \,dr\,
J_{\ell-\ell'}(k\,r\sin\theta)
\,\ILag{n}{s}{\rho}\,\ILag{n'}{s'}{\rho}
= \ILag{n}{n'}{\bar x}\,\ILag{s'}{s'}{\bar x}
\eeqa
where we have used $\rho = \gamma r^2$ from \eqref{gamma_rho}, and defined
\beq
\label{eq:xbar}
\bar x = \frac{|\vec q|^2 \sin^2\theta_X}{4\gamma}
\eeq
In the transition rate, using the Lagurre-Gaussian's completeness relations, the only $s'$ dependence then reads
\beq
\label{eq:ILag_completeness}
\sum_{s'=0}^{\infty} \left( \ILag{s}{s'}{\bar x}\right)^2 = 1
\eeq

The sum, $\sum_{n'=0}^n$, over final state fermion Landau-levels can be exchanged by an integral over the difference of initial and final levels, $\nu=n-n'$,
\beq
\label{eq:nu_int}
\sum_{n'=0}^n \to \int_{0}^{n} d\nu
\eeq
In order to ease calculations, we choose a reference-frame in which $k_3=0$.
Results for non-zero, $k_3$ can be obtained by boosting along the 3-direction.
The initial and final state energies then read
\beq
\label{eq:energies}
E_i = \sqrt{m^2 +4n\gamma},
\qquad
E_f = \sqrt{m^2 + 4(n-\nu)\gamma + |\vec q|^2\cos\theta_X},
\qquad
E_q = \sqrt{|\vec q|^2 + m_X^2}
\eeq
and also
\beq
\label{eq:k3p}
k'_3 = -|\vec q|\cos\theta_X
\eeq

The $\nu$-integral can be performed over the $\delta$-function using
\beq
\tfrac{d}{d\nu}\left(E_i - E_f - E_q\right)  = \frac{2\gamma}{E_F}
\eeq
and using the solution
\beq
\label{eq:nu_sol}
\nu = \frac{2E_i E_q - m_X^2 - |\vec q|^2\sin\theta_X^2}{4\gamma} \,
\eeq
which, in turn, implies
\beq
m_X < E_q < E_i - m,
\qquad
0 < |\vec q| < \sqrt{(E_i - m)^2 - m_X^2} \equiv \qmax
\eeq
Finally, the transition rate reads
\beq
\label{eq:trans_rate}
\boxed{
\Gamma_X^{u\to u'} = 
\frac{1}{2\pi}\int_{-1}^{1} d(\cos\theta_X)
\int_0^{\qmax} d|\vec q|\,\frac{|\vec q|^2}{2E_q}\frac{E_f}{2\gamma}
\,
\left(\frac{|M_{uu'}|^2}{2E_i 2E_f}\right)
}
\eeq 
where \eqref{JII}, \ref{eq:ILag_completeness}, \ref{eq:energies}, \ref{eq:k3p}, and \ref{eq:nu_sol}
are assumed.

It is useful to perform a variable change which rescales the momentum integral to the $[0,\infty)$ domain,
\beq
y = \frac{1}{\xi_0} \frac{|\vec q|/E_i}{1-|\vec q|/\qmax}
\qquad
\Leftrightarrow
\qquad
|\vec q| = E_i \frac{ \xi_0\,y}{1+\xi_0\, y \,E_i/\qmax }
\eeq
where $\xi_0$, defined in \eqref{xi0_def}, is used, foreseeing a small parameter expansion.
With $\xi_0 \ll 1$, and the small parameter expansions in next subsections, one finds several useful approximate expression which are listed in \appref{useful}.

When these results are put in correctly, it turns out that the only meaningful effect of the $m_X\ne0$ is in the (phase-space) integrals over Bessel-K functions, while other effects in $m_X/m$, and $m_X^2/E_i^2$ are sub-leading.
In the $y,~\cos\theta_X$ variables, the $X$-emission rate can then be written as

\beq
\boxed{
\Gamma_X^{u\to u'} = 
\frac{27}{32\pi(m/E_i)^9 E_i\, \rho^2\,\xi_0}\int_{0}^{\infty}dy\,
\frac{y}{\left(1+\xi_0\,y \right)^4}
\int_{-1}^{1}
 d(\cos\theta_X)
\,
\left(\frac{\left|M_{uu'}^{\rm{LO}}\right|^2}{4E_i E_f}\right)
}
\eeq
where $M_{uu'}^{\rm{LO}}$ stands for the leading order expansions described in the subsequent sub-sections\footnote{
for similar considerations, we approximate $E_i / \qmax\approx 1$
Note that we have kept the $E_i E_f$ factor in the ratio, due to our ``particle-physics'' normalization of the wave-functions.
}

\subsection{Small Parameter Expansions of the Matrix Element}
\label{sec:small_param_exp}

In a planner high-energy storage-ring, the particles are ultra-relativistic, $m\ll E_i$, and the emission of radiation therefore occurs close to the plane of the ring, at small $|\cos\theta_X|$.
The kinematic variables can be expanded in
\beq
\label{eq:eps_0_eps}
\epsilon_0 = 1-\beta^2 = \frac{m^2}{E_i^2},
\qquad
\epsilon  = 1-\beta^2\sin^2\theta_X
\eeq
which are the small parameters for the synchrotron case ($\cos^2\theta_X = \frac{\epsilon-\epsilon_0}{1-\epsilon_0}$).
With the emission of a light $X$ particle, there is another small parameter
\beq
\epsilon_x = \frac{m_X^2}{E_q^2} \approx \frac{m_X^2}{|\vec q|^2}
\eeq
where $E_q = \sqrt{|\vec q|^2 + m_X^2}$, and $|\vec q|$ are the energy and momentum of the $X$ particle, which is given as a solution to the equation
\beq
|\vec q|^2 = (E_i  - Ef)^2 - m_X^2 = \left(\sqrt{m^2+4\gamma n} - \sqrt{m^2+4\gamma n + |\vec q|^2\cos\theta^2_X}\right)^2 - m_X^2
\eeq
The solution, and additional kinematic variables can be expanded in $\epsilon_0,\,\epsilon,\, \epsilon_X$ to give
\beqa
|\vec q| &\approx& \sqrt{4\gamma}\left(\sqrt{n} - \sqrt{n'} \right)
\left( 
1 - \epsilon\frac{ \sqrt{n} - \sqrt{n'} }{2\sqrt{n'} } - \frac{\epsilon_0}{2} - \frac{\epsilon_X}{2} 
\right)
\\
1 - \frac{\bar x}{x_0} &\approx& 1- \frac{|\vec q|^2\sin^2\theta_X}{4\gamma\left( \sqrt{n} - \sqrt{n'} \right)^2}
\approx \sqrt\frac{n}{n'}\epsilon + \epsilon_X
\eeqa
where $x_0$ is defined below in \eqref{turning_pts}.
Similarly,
\beqa
y &=& \frac{1}{\xi_0} \frac{|\vec q|/E_i}{1-|\vec q|/\qmax}
\approx
\tfrac43
\epsilon_0^{3/2}
n
\sqrt\frac{x_0}{n'}
\left(
1-
\frac{\sqrt{n x_0} }{n'}\frac{\epsilon}{2}
-
\left(
1+\sqrt\frac{x_0}{n}+\frac{x_0}{n}
\right)
\frac{\epsilon_X}{2}
+
\frac{x_0}{n'}\epsilon_0
\right)
\eeqa
and
\beq
\sqrt\frac{n}{n'} \approx \frac{E_i}{E_f} = (1+\xi_0\, y E_i / \qmax)
\left(
1+ \frac{m_X^2}{2E_i^2}\frac{1+2\xi_0\, y}{\xi_0\, y}
\right)
\eeq

\subsection{WKB Approximation of The Lagurre Gaussians $\ILag{n}{n'}{x}$}
\label{sec:WKB}
In~\cite{Sokolov:1986nk} an approximate form is derived for the Lagurre Gaussian functions, $\ILag{n}{n'}{x}$.
Here we summarize the relevant results, and generalize them to the massive case.
The differential equation for $\ILag{n}{n'}{x}$ is
\beq
\label{eq:ilag_schrodinger}
\frac{d^2}{dx^2}\left(x^{1/2} \ILag{n}{n'}{x}\right)
-
f(x)x^{1/2} \ILag{n}{n'}{x}
\eeq
where $f(x) = 0$ is solved by
\beq
\label{eq:turning_pts}
\begin{matrix}
x_0 \\
x_0'
\end{matrix}
\bigg\}
=
(n+n'+1)\mp\left(4nn' + 2 (n+n'+1)\right)^{1/2} \approx \left(n^{1/2}\mp n'^{1/2}\right)^{1/2}
\eeq
and the approximation is for $n\gg1$, $n'\gg1$ (as would be appropriate when the Landau-level describes a macroscopic classical trajectory in the storage-ring).

\Eqref{ilag_schrodinger} is a Schr{\"o}dinger-like equation for $\left(x^{1/2} \ILag{n}{n'}{x}\right)$. An approximate solution can be obtained close to the turning points of the potential $\{x_0,\,x_0'\}$ using the WKB method. Sokolov \& Ternov~\cite{Sokolov:1986nk} obtain
\beq
\begin{cases}
\ILag{n}{n}{x} = \frac{1}{\pi\sqrt3 }\left(1-\frac{x}{x_0}\right)^{1/2}K_{1/3}(z) \\
\ILag{n}{n}{x}' = \frac{(nn')^{1/4}}{\pi\sqrt{3x_0}}\left(1-\frac{x}{x_0}\right)K_{2/3}(z) 
\end{cases}
\qquad
\text{for~~}
x=x_0 + 0^+
\eeq
where the $0^+$ notation indicates that $x$ is close to, but larger than $x_0$, and 
\beq
z = \frac{2}{3}\left(x_0^2 nn'\right)^{1/4}\left(1-\frac{x}{x_0}\right)^{3/2}
\eeq

\beqa
\ILag{n}{n'-1}{x} &=&
\sqrt\frac{x}{n'}\left(
\frac{n-n'-x}{2x}\ILag{n}{n}{x}
-
\ILag{n}{n'}{x}'
\right)
\\
\ILag{n-1}{n'}{x} &=& 
\sqrt\frac{x}{n}\left(
\frac{n-n'+x}{2x}\ILag{n}{n}{x}
+
\ILag{n}{n'}{x}'
\right)
\\
\ILag{n-1}{n'-1}{x} &=&
\frac{x}{\sqrt{n\,n'}}\left(
\frac{n+n'+x}{2x}\ILag{n}{n}{x}
-
\ILag{n}{n'}{x}'
\right)
\eeqa

Using the small parameter approximations of \appref{small_param_exp}, we find
\beqa
1- \frac{x}{x_0} &\approx& 
\sqrt\frac{n}{n'}\left(\epsilon+\frac{n'}{n}\epsilon_X\right)
=
\sqrt\frac{n}{n'}\left(\epsilon + \frac{m_X^2}{E_i^2}\frac{1+\xi_0\, y}{(\xi_0\, y)^2}\right)
\\
\label{eq:z_approx}
z &\approx& \tfrac12 y \epsilon_0^{3/2}\left(\epsilon+\frac{n'}{n}\epsilon_X\right)^{3/2}
=
\tfrac12 y \epsilon_0^{3/2}
\left(
\epsilon+\frac{m_X^2}{E_i^2}\frac{1+\xi_0\, y}{(\xi_0\, y)^2}\right)^{3/2}
\eeqa
so that
\beqa
\begin{pmatrix}
\ILag{n}{n'}{\bar x} \\
\ILag{n}{n'-1}{\bar x} \\
\ILag{n-1}{n'}{\bar x} \\
\ILag{n-1}{n'-1}{\bar x} 
\end{pmatrix}
&=&
\frac{\sqrt{1+\xi_0\, y E_i / \qmax} }{\pi\sqrt3}
\left(
1+ \frac{m_X^2}{4E_i^2}\frac{1+2\xi_0\, y}{\xi_0\, y}
\right)
\left(
\epsilon+\frac{m_X^2}{E_i^2}\frac{1+\xi_0\, y}{(\xi_0\, y)^2}
\right)^{1/2}
\nn\\
&&\left(
K_{1/3}(z) +
\begin{pmatrix}
0 \\
-(1+\xi_0\, y)\left(
1+ \frac{m_X^2}{4E_i^2}\frac{1+2\xi_0\, y}{\xi_0\, y}
\right)
\\
1 \\
-\xi_0\, y - (1+\xi_0\, y)\frac{m_X^2}{4E_i^2}\frac{1+2\xi_0\, y}{\xi_0\, y}
\end{pmatrix}
\left(
\epsilon+\frac{m_X^2}{E_i^2}\frac{1+\xi_0\, y}{(\xi_0\, y)^2}
\right)^{1/2}
K_{2/3}(z)
\right)
\nn\\
\eeqa

\section{Integral Evaluation}
\label{sec:int_eval}

The Bessel-K functions we use as a result of the WKB approximation (\appref{WKB}) lead to angular integrals of the form
\beq
\label{eq:ang_int}
I\left(a,b,c,d\right)
\equiv
\int_{-1}^{1}\,d\cos\theta\,
\left(
\epsilon+\frac{m_X^2}{E_i^2}\frac{1+\xi_0\, y}{(\xi_0\, y)^2}
\right)^{a}
\cos^d\theta\,
K_{b}(z)K_{c}(z)
\eeq
where $a\in\{\tfrac12,\,1,\, \tfrac32\}$, $b,\, c \in \{\tfrac13,\,\tfrac23 \}$, $d$ is a non-negative even integer, and $z$ is given in
\eqref{z_approx}, and $\epsilon = \left(1 - \epsilon_0\right)\cos^2\theta + \epsilon_0$ via
\eqref{eps_0_eps}.
The integrand in \eqref{ang_int} is an even function of $\cos\theta$ on the $[-1,1]$ domain.
In addition, the Bessel-K functions decay quickly for large arguments.
The integral can then be approximated by evaluating the integral in the domain $[0,\infty)$, and multiplying by $2$.
This class of integrals has been evaluated in~\cite{Aspnes:1966zz}, with the results given as a recursion relation (see also~\cite{Jackson:1975qi}, and~\cite{Sokolov:1986nk}). We apply the prescription and obtain results for the case of massive particles. We define
\beqa
\zeta &\equiv& \left(1 + \frac{1 + \xi_0\, y}{(\xi_0\, y )^2}\, \frac{m_{X}^2}{m^2}\right)
\eeqa
and find
\beqa
I(1,\tfrac13,\tfrac13,0)
&=&
\frac{2\pi}{\sqrt3y}\epsilon_0^{3/2}
\int_{y\, \zeta^{3/2}}^\infty K_{1/3}(x)\,dx
\\
I(1,\tfrac13,\tfrac13,2)
&=&
\frac{\pi}{\sqrt3y}\epsilon_0^{5/2}\zeta\left(
\int_{y\, \zeta^{3/2}}^\infty K_{5/3}(x)\, dx
-
K_{2/3}(y\, \zeta^{3/2})
\right)
\\
I(1,\tfrac13,\tfrac13,4)
&=&
\frac{\pi}{2\sqrt3y^2}\epsilon_0^{7/2}\zeta^{1/2}\left(
K_{1/3}(y\, \zeta^{3/2})
-
\frac{3y}{2}\zeta^{3/2}
\left(
\int_{y\, \zeta^{3/2}}^\infty K_{5/3}(x)\, dx
-
K_{2/3}(y\, \zeta^{3/2})
\right)
\right)
\\
I(2,\tfrac23,\tfrac23,0)
&=&
\frac{\pi}{\sqrt3y}\epsilon_0^{5/2}\zeta
\left(
\int_{y\, \zeta^{3/2}}^\infty K_{5/3}(x)\, dx
+
K_{2/3}(y\, \zeta^{3/2})
\right)
\\
I(2,\tfrac23,\tfrac23,2)
&=&
\frac{\pi}{6\sqrt3 y^2}\epsilon_0^{7/2}\left(
5\zeta^{1/2}K_{1/3}(y\, \zeta^{3/2})
+
\frac{3y}{2}\zeta^{2}
\left(
\int_{y\, \zeta^{3/2}}^\infty K_{1/3}(x)\, dx
-
K_{2/3}(y\, \zeta^{3/2})
\right)
\right)
\\
I(\tfrac32,\tfrac13,\tfrac23,0)
&=&
\frac{2\pi}{\sqrt3 y}\epsilon_0^{2}\zeta^{1/2}K_{1/3}(y\, \zeta^{3/2})
\\
I(\tfrac32,\tfrac13,\tfrac23,2)
&=&
\frac{2\pi}{3\sqrt3 y^2}\epsilon_0^{3}\int_{y\, \zeta^{3/2}}^\infty K_{1/3}(x)\, dx
\\
\eeqa
In presenting our results, we follow~\cite{Sokolov:1986nk} and use the relation
\beq
K_{1/3}(\alpha) + K_{5/3}(\alpha) = -2\frac{d}{d\alpha}K_{2/3}(\alpha)
\eeq


\section{Useful Expansions of Kinematic Variables}
\label{sec:useful}

Some useful expansions
\beqa
|\vec q| &=& E_i \frac{ \xi_0\,y}{1+\xi_0\, y \,E_i/\qmax }
\\
E_f &=& E_i \frac{ 1}{1+\xi_0\, y \,E_i/\qmax }\left( 
1 - \frac{m_X^2}{2E_i^2}\frac{1+2\,\xi_0\,y}{\xi_0\,y}
\right)
\\ 
\frac{k'_3}{E_f} &=& - \xi_0\, y\,\cos\theta_X\left( 1 + \frac{m_X^2}{2E_i^2} \right)
\\ 
\sqrt{1- \frac{{k'}^2_3}{E^2_f}}&=& 1-\tfrac12(\xi_0\,y)^2\cos\theta_X^2\left( 1 + \frac{m_X^2}{2E_i^2} \right)
\\ 
E_f^2 - {k'}^2_3 &=&  \left(\frac{E_i}{1+\xi_0\, y \,E_i/\qmax }\right)^2
\left( 
1 - \frac{m_X^2}{E_i^2}\frac{1+2\,\xi_0\,y}{\xi_0\,y} - (\xi_0\,y)^2\cos\theta_X^2
\right)
\\ 
\frac{E_i}{\qmax} &\approx& 1 + \frac{m}{E_i} + \frac{m_X^2}{2E_i^2}
\\ 
\eeqa

\section{Spin-1 Polarization Vectors}
\label{sec:spin1_pols}
The transverse and longitudinal polarization vectors for spin-1 are
\beqa
&&\left(\varepsilon^T\right)^{\chi=\pm}_\mu = \tfrac{1}{\sqrt2}
\begin{pmatrix}
0, & 1, & i\,\chi\cos\theta_X, & -i\,\chi\sin\theta_X
\end{pmatrix}
\\
&&\left(\varepsilon^L\right)_\mu = 
\begin{pmatrix}
\frac{|\vec q|}{m_X}, & 0, & \frac{E_q}{m_X}\sin\theta_X, & \frac{E_q}{m_X}\cos\theta_X 
\end{pmatrix}
\eeqa


\section{Full Results of Spin-Flip Transition Rates}
\label{sec:full_results}

The spin-flip transition-rate by $X$-emission,
\beq
\Gamma^{\rm sf}_{X}(u) \equiv
\Gamma_H(e^-(u) \to e^-(-u) \gamma) \, .
\eeq
depends on the initial spin state of the fermion, $u=+1,\, -1$ for ''up`` and ``down'' respectively.
The emitting state is an electron of mass $m$, and energy $E_i$, and the bending radius is $R$.
We define
\beqa
y &=& \frac{1}{\xi_0} \frac{|\vec q|/E_i}{1-|\vec q|/\qmax} \\
\zeta &\equiv& \left(1 + \frac{1 + \xi_0\, y}{(\xi_0\, y )^2}\, \frac{m_{X}^2}{m^2}\right)
\\
{\cal K} &\equiv& \int_{y\, \zeta^{3/2}}^\infty dx\, K_{5/3}(x), \qquad
\overset{--}{{\cal K}} \equiv \int_{y\, \zeta^{3/2}}^\infty dx\, K_{1/3}(x), \qquad 
\\
K_{1/3} &\equiv& K_{1/3}(y\,\zeta^{3/2}), \qquad
K_{2/3} \equiv K_{2/3}(y\,\zeta^{3/2})
\eeqa

The rates for the various $X$ are given by
\beqa
\frac{d\,\Gamma^{\rm sf}_{V_T}}{d\,y} &=&
g_V^2\,\frac{\sqrt3 E_i}{2^5 \pi^2 m \rho}
\left(
\frac{1}{(1 + \xi_0\, y)^2}\frac{m_{V}^2}{m^2}
{\cal K}
+
\frac{2(\xi_0\, y)^2}{(1 + \xi_0\, y)^3}
\left(
K_{2/3}
+
u\,\zeta^{1/2} K_{1/3}\,
\right)
\right)
\\
\frac{d\,\Gamma^{\rm sf}_{V_L}}{d\,y} &=&
g_V^2\,\frac{\sqrt3 E_i}{2^5 \pi^2 m \rho}
\left(\frac{m^2 m_V^2}{16 E_i^4}\right)
\frac{(\xi_0\, y)^2}{1 + \xi_0\, y}
\Bigg(
\left(
2\zeta
-\frac{\zeta^{1/2}}{2}
-\frac{8u}{3y}
\right)
{\cal K}
+
\left(
-2\zeta
+\frac{\zeta^{1/2}}{2}
+\frac{16u}{3y}
\right)
K_{2/3}
+
\frac{5\zeta^{1/2}}{3y}K_{1/3}
\Bigg)
\nn\\\\
\frac{d\,\Gamma^{\rm sf}_{A_T}}{d\,y} &=&
g_A^2\,\frac{\sqrt3 E_i}{2^5 \pi^2 m \rho}
\Bigg(
\frac{(2+\xi_0\, y)^2}{(1 + \xi_0\, y)^2(\xi_0\, y)^2}\frac{m_{A}^2}{m^2}
{\cal K}
+
\frac{2(2+\xi_0\, y)^2}{(1 + \xi_0\, y)^3}
\bigg(
K_{2/3}
+
u\, \zeta^{1/2} K_{1/3}\,
\bigg)
\Bigg)
\\
\frac{d\,\Gamma^{\rm sf}_{A_L}}{d\,y} &=&
g_A^2\,\frac{\sqrt3 E_i}{2^5 \pi^2 m \rho}
\Bigg(
\left(\frac{m^2}{m_{A}^2}\frac{(\xi_0\, y)^2}{(1 + \xi_0\, y)^3}\right)
\left(
\left(\zeta-2\right)
{\cal K}
+
(4+\zeta)K_{2/3}
+
4u\, \zeta^{1/2} K_{1/3}\,
\right)
\nn\\
+
&&2\frac{m_A^2}{m^2}
\frac{1}{ (\xi_0\, y)^2(1 + \xi_0\, y) }\overset{--}{{\cal K}}
- \frac{4 u \zeta^{1/2} }{(1+\xi_0\, y)^2}K_{1/3}~~
\Bigg)
\\
\frac{d\,\Gamma^{\rm sf}_S}{d\,y} &=&
g_S^2\,\frac{\sqrt3 E_i}{2^5 \pi^2 m \rho}
\left(
8 \frac{(\xi_0\, y)^2}{(1 + \xi_0\, y)^3} \zeta
\right)
\left(
{\cal K}
+
K_{2/3}
\right)
\\
\frac{d\,\Gamma^{\rm sf}_a}{d\,y} &=&
g_a^2\,\frac{\sqrt3 E_i}{2^5 \pi^2 m \rho}
\left(\frac{(\xi_0\, y)^2}{(1 + \xi_0\, y)^3}\right)
\Bigg(
\left(\zeta-2\right)
{\cal K}
+
(4+\zeta)K_{2/3}
+
4u\, \zeta^{1/2} K_{1/3}\,
\Bigg)
\eeqa
where for the massive vector, and axial-vector cases we distinguish between the transverse, and longitudinal modes, denoting them by $_T$, and $_L$ respectively.

The result of \secref{spin_flip_rates} are obtained by taking the limit $\frac{m_X}{m \,\xi_0}\to0$, (or $\zeta\to1$). In this limit, the order of the $y$ and $x$ integrals can be interchanged with explicit expression for the limits, and the integrals are straightforward to preform after an expansion in $\xi_0$.

\end{document}